\documentclass[english]{article}
\usepackage[T1]{fontenc}
\usepackage[latin1]{inputenc}
\usepackage{geometry}
\usepackage{subfigure}
\usepackage{amsmath}
\usepackage{graphicx}
\usepackage{amssymb}
\usepackage[numbers]{natbib}

\makeatletter
 \newcommand{\lyxaddress}[1]{
   \par {\raggedright #1
   \vspace{1.4em}
   \noindent\par}
 }

\usepackage{subfigure}
\usepackage{graphicx}

\usepackage{babel}
\makeatother
\begin{document}

\title{Comparative analysis of some models of mixed-substrate microbial
growth}

\author{Atul Narang%
\thanks{Email: \texttt{narang@che.ufl.edu}%
}}

\maketitle

\lyxaddress{Department of Chemical Engineering, University of Florida, Gainesville,
FL~32611-6005. }

\noindent \begin{flushleft}Keywords: Mathematical model, mixed substrate
growth, diauxic growth, lac operon, Lotka-Volterra model.\end{flushleft}

\begin{abstract}
Mixed-substrate microbial growth is among the most intensely studied
systems in molecular microbiology. Several mathematical models have
been developed to account for the genetic regulation of such systems,
especially those resulting in diauxic growth. In this work, we compare
the dynamics of three such models (Narang, Biotech.~Bioeng., 59,
116, 1998; Thattai \& Shraiman, Biophys.~J, 85, 744, 2003; Brandt
et al, Water Research, 38, 1004, 2004). We show that these models
are dynamically similar --- the initial motion of the inducible enzymes
in all the models is described by Lotka-Volterra equations for competing
species. In particular, the prediction of diauxic growth corresponds
to {}``extinction'' of one of the enzymes during the first few hours
of growth. The dynamic similarity occurs because in all the models,
the inducible enzymes possess properties characteristic of competing
species: Their synthesis is autocatalytic, and they inhibit each other.
Despite this dynamic similarity, the models vary with respect to the
range of dynamics captured. The Brandt et al model captures only the
diauxic growth pattern, whereas the remaining two models capture both
diauxic and non-diauxic growth patterns. The models also differ with
respect to the mechanisms that generate the mutual inhibition between
the enzymes. In the Narang model, the mutual inhibition occurs because
the enzymes for each substrate enhance the dilution of the enzymes
for the other substrate. In the Thattai \& Shraiman model, the mutual
inhibition is entirely due to competition for the phosphoryl groups.
Elements of all the models appear to be necessary for quantitative
agreement with data.
\end{abstract}

\section{Introduction\label{s:Introduction}}

\begin{figure}
\begin{center}\subfigure[]{\includegraphics[%
  width=7cm,
  height=5cm]{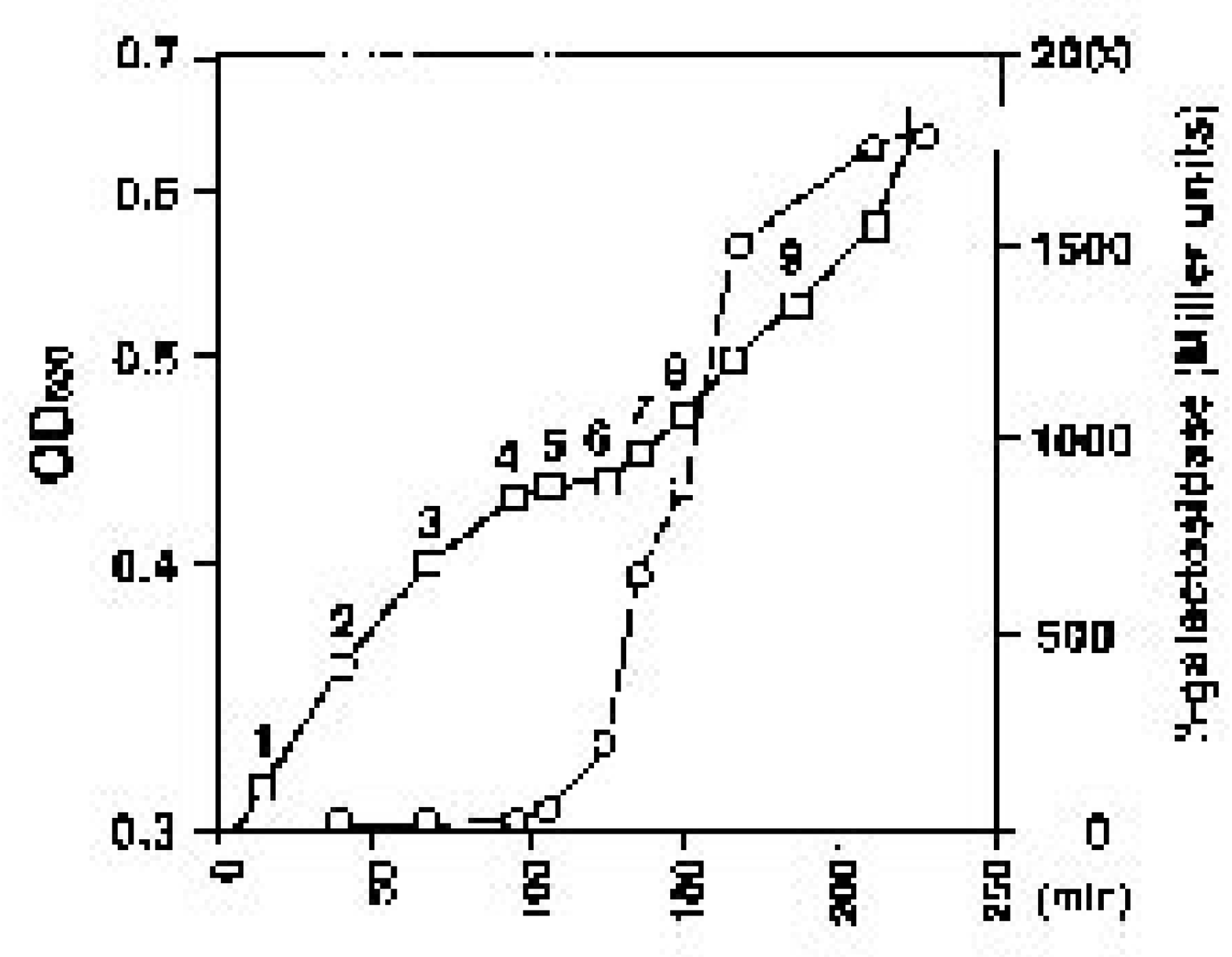}}\hspace{0.1in}\subfigure[]{\includegraphics[%
  width=7cm,
  height=5cm]{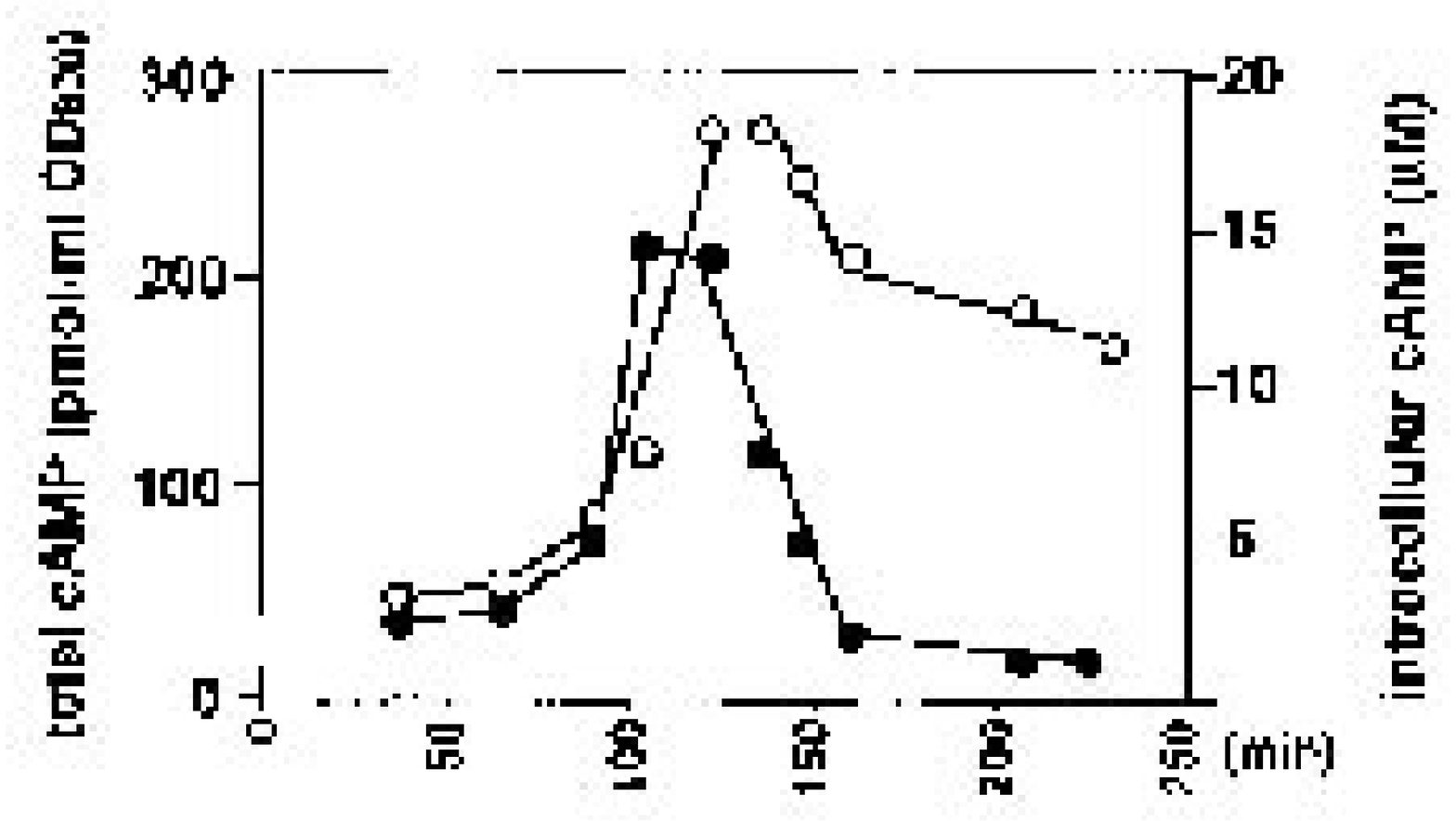}}\end{center}

\caption{\label{f:DiauxicGrowth}Diauxic growth of \emph{E. coli} on a mixture
of glucose and lactose (from~\citep{inada96}): (a) The optical density~($\square$)
shows two exponential growth phases separated by a diauxic lag ($60\lesssim t\lesssim160$~min).
The levels of $\beta$-galactosidase~($\bigcirc$), a peripheral
enzyme for lactose, remain low until the beginning of the diauxic
lag. (b)~Evolution of the intracellular cAMP levels during the experiment
shown in (a). The intracellular cAMP levels~($\bullet$) during the
first phase of exponential growth on glucose ($t\lesssim60$~min)
are similar to the intracellular cAMP levels during the second phase
of exponential growth on lactose ($t\gtrsim160$~min).}
\end{figure}

When microbial cells are grown in a batch culture containing a mixture
of two carbon sources, they often exhibit \emph{diauxic} growth, which
is characterized by the appearance two exponential growth phases separated
by a lag phase called \emph{diauxic lag}~\citep{monod47}. The most
well-known example of this phenomenon is the batch growth of \emph{E.
coli} on a mixture of glucose and lactose (Figure~\ref{f:DiauxicGrowth}a).
Early studies by Monod showed that in this case, the two exponential
growth phases reflect the sequential consumption of glucose and lactose~\citep{monod1}.
Moreover, only glucose is consumed in the first exponential growth
phase because the synthesis of the \emph{peripheral} enzymes for lactose
(the enzymes that catalyze the transport and peripheral catabolism
of lactose) is somehow abolished in the presence of glucose. During
this period of preferential growth on glucose, the peripheral enzymes
for lactose are diluted to very small levels: 6--7 generations of
growth on glucose reduce the enzyme levels to $\sim1$\% of their
initial values. Thus, the diauxic lag reflects the time required for
the cells to build up the peripheral enzymes for lactose to sufficiently
high levels. After the diauxic lag, one observes the second exponential
phase corresponding to consumption of lactose.

The key to the resolution of the glucose-lactose diauxie is clearly
the molecular mechanism by which the synthesis of lactose-specific
enzymes is abolished in the presence of glucose. The first inroads
into this problem were made by Monod and coworkers who discovered
the mechanism for synthesis (induction) of the lactose-specific enzymes
in the presence of lactose~\citep{jacob61}. It was shown that the
genes corresponding to the peripheral enzymes for lactose are contiguous
on the DNA, an arrangement referred to as the \emph{lac} operon. In
the absence of lactose, transcription of the \emph{lac} operon is
prevented by a repressor molecule, called the \emph{lac} repressor,
which is bound to a specific site on the \emph{lac} operon. In the
presence of lactose, transcription of the \emph{lac} operon is triggered
because allolactose, a product of $\beta$-galactosidase, sequesters
the repressor from the operon, thus liberating it for transcription.%
\footnote{A similar mechanism serves to induce the genes for glucose transport~\citep[Figure~4]{plumbridge03}.
In the absence of glucose, transcription of the \emph{ptsG} gene,
which codes for the transport enzyme, IIBC$^{{\rm glc}}$, is inhibited
because a repressor called Mlc is bound to a regulatory site \emph{}on
the gene. Upon entry of glucose, IIBC$^{{\rm glc}}$ is dephosphorylated.
Dephosphorylated IIBC$^{{\rm glc}}$ sequesters Mlc away from the
regulatory site on \emph{ptsG}, thus liberating the gene for transcription.%
}

Given this mechanism for induction of the lactose-specific enzymes,
it seems plausible to hypothesize that the glucose-lactose diauxie
occurs because transcription of the \emph{lac} operon is somehow abolished
in the presence of glucose. These mechanisms are not fully understood~\citep{stulke99}.
Until recently, there were two models for inhibition of \emph{lac}
transcription in the presence of glucose

\begin{enumerate}
\item \emph{cAMP activation:} This model postulates that a complex consisting
of cyclic AMP (cAMP) and catabolite repression protein (CRP) must
bind to a specific site on the \emph{lac} operon before it can be
transcribed. When glucose is added to a culture growing on lactose,
the cAMP levels somehow decrease, which reduces the binding of the
cAMP-CRP complex to the \emph{lac} operon, thus inhibiting its transcription
rate.
\item \emph{Inducer exclusion:} According to this model, enzyme IIA$^{{\rm glc}}$,
a peripheral enzyme for glucose, is dephosphorylated in the presence
of glucose. The dephosphorylated enzyme IIA$^{{\rm glc}}$ inhibits
lactose uptake by binding to the lactose permease, the transport enzyme
for lactose. This reduces the intracellular concentration of allolactose,
and hence, the transcription rate of the \emph{lac} operon.
\end{enumerate}
Experiments by Aiba and coworkers have shown that the cAMP activation
model is not tenable~\citep{inada96}. The cAMP levels are the same
during growth on glucose and lactose (Figure~\ref{f:DiauxicGrowth}b).
Moreover, the \emph{lac} operon is not transcribed even if cAMP is
added to a culture growing on glucose and lactose. It is now believed
that inducer exclusion alone is responsible for inhibiting \emph{lac}
transcription. But in \emph{E. coli} ML30, the activity of lactose
permease is inhibited no more than $\sim$40\% at saturating concentrations
of glucose~\citep[Table 2]{cohn59a}. Likewise, Saier and coworkers,
who discovered inducer exclusion in \emph{S. typhimurium}, found that
inducer exclusion by glucose inhibits the synthesis of the peripheral
enzymes for melibiose, glycerol, maltose, and lactose by 10--50\%~\citep[Figures~1--2]{Saier1972}.
This partial inhibition by inducer exclusion cannot explain the almost
complete inhibition of the genes for the {}``less preferred'' substrates.

\begin{figure}
\begin{center}\subfigure[]{\includegraphics[%
  width=7cm,
  height=5cm]{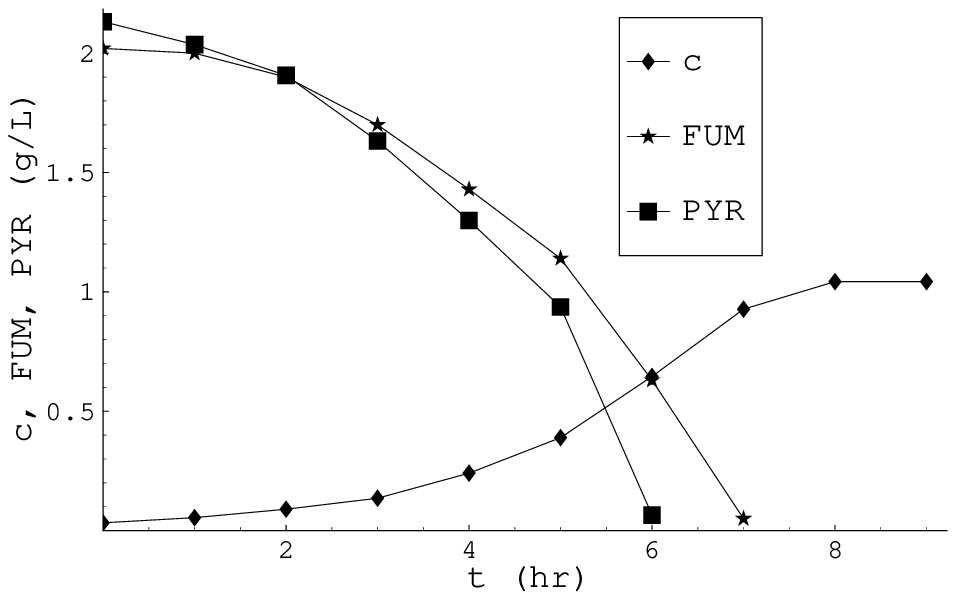}}\hspace{0.1in}\subfigure[]{\includegraphics[%
  width=7cm,
  height=5cm]{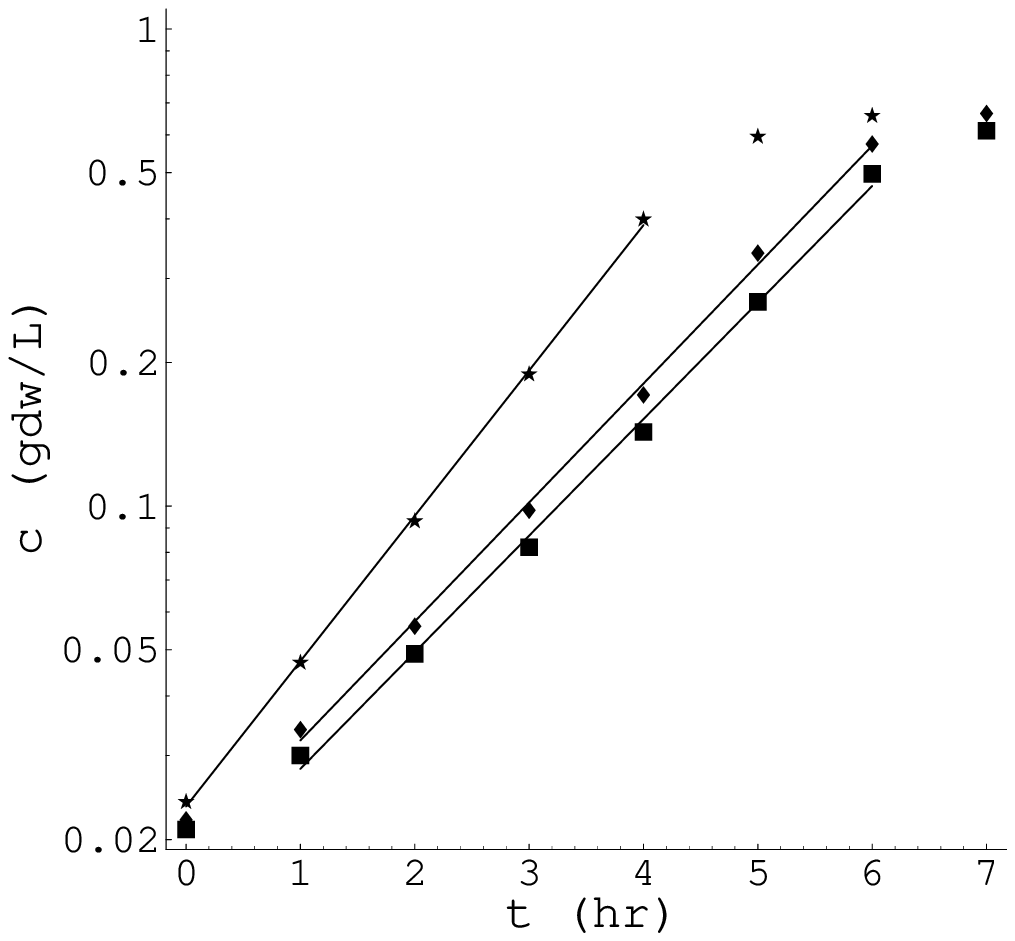}}\end{center}

\caption{\label{f:NondiauxicGrowth}Nondiauxic growth of \emph{E. coli} (from~\citep{narang97a}):
(a) Simultaneous substrate utilization during batch growth of \emph{E.
coli} K12 on a mixture of fumarate (FUM) and pyruvate (PYR). The cell
density is denoted by $c$ (gdw/L) . This growth pattern is observed
with several pairs of organic acids~\citep{narang97a}. (b) Growth
pattern dependent on the history of the inoculum. When the inoculum
is grown on glucose ($\ast$), the specific growth rate on a mixture
of glucose and pyruvate is 0.74~1/hr. When the inoculum is grown
on pyruvate ($\blacksquare,\blacklozenge$), the specific growth rate
on the same mixture is 0.56~1/hr.}
\end{figure}

Although the diauxie has dominated the literature on mixed-substrate
growth, there is ample evidence of non-diauxic growth patterns. This
was already evident from Monod's early studies in which he classified
his mixed-substrate data into two categories~\citep{monod1,monod47}.
Growth on a particular mixture was called diauxic if the growth curve
showed the diauxic lag, and \emph{normal} if it showed no such lag.
Yet, the phenomenon of normal growth was virtually ignored until recently.
In the last few years, several studies have shown that both substrates
can be consumed simultaneously. Figure~\ref{f:NondiauxicGrowth}a
shows, for instance, that \emph{E. coli} consumes fumarate and pyruvate
simultaneously during batch growth. Egli has summarized all known
examples of simultaneous substrate utilization in a comprehensive
review article~\citep{egli95}. He notes that, in general, simultaneous
substrate utilization is observed when both substrates support low-to-medium
specific growth rates, and diauxic growth occurs when one of the substrates
supports a specific growth rate that is substantially higher than
the specific growth rate on the other substrate. In addition to simultaneous
substrate utilization, there is some evidence that the substrate utilization
pattern can depend on the history of the inoculum, one example of
which is shown in Figure~\ref{f:NondiauxicGrowth}b (see~\citep{narang98b}
for other examples).

The phenomenon of mixed-substrate growth is of fundamental importance
in molecular biology as a paradigm of the mechanism by which the expression
of DNA is controlled. It also has profound implications for several
large-scale biotechnological processes.%
\footnote{The large-scale production of chemicals, such as bioethanol and biopolymers,
is economically feasible only if they are derived from cheap lignocellulosic
feedstocks~\citep{ingram99}. The pretreatment of these feedstocks
yields a mixture of hexoses (primarily, glucose) and pentoses (primarily,
xylose). The cells that ferment these sugars to useful products typically
exhibit diauxic growth with preferential consumption of hexoses.%
} This has spurred the development of several mechanistic models of
mixed-substrate growth. Some of these models are inspired by the detailed,
but constantly evolving, knowledge of the molecular mechanism for
the glucose-lactose diauxie~\citep{Santillan2004,vandedem73,wong97}.
The other models appeal to the fact that the phenomenon of diauxic
growth is ubiquitous --- it has been observed in diverse microbial
species on many pairs of \emph{substitutable} substrates (i.e., substrates
that satisfy the same nutrient requirements) including pairs of carbon~\citep{egli95,harder82,kovarova98},
nitrogen~\citep{Neidhardt1957}, and phosphorus~\citep{Daughton1979}
sources, and even among pairs of electron acceptors~\citep{Liu1998}.
Thus, it is conceivable there exist some general mechanisms driving
the dynamics of mixed-substrate growth. These general models, which
abstract features common to many, if not all, mixed-substrate systems,
include the cybernetic model~\citep{kompala84,ramakrishna96} and
several kinetic models~\citep{Brandt2003,narang97c,Thattai2003}.

The original cybernetic model, which was analyzed in~\citep{narang97b},
cannot capture nondiauxic growth patterns. In this work, we compare
the general kinetic models developed by Narang et al~\citep{narang98b,narang97c},
Brandt et al~\citep{Brandt2003}, and Thattai \& Shraiman~\citep{Thattai2003}.
Hereafter, we shall refer to these as the N-, B- and T-models, respectively.
We show that

\begin{enumerate}
\item All the models are similar inasmuch as they exhibit the same general
class of dynamics. More precisely, the equations describing the initial
evolution of the peripheral enzymes are special cases of the generalized
Lotka-Volterra model for competing species. This similarity arises
because all the models possess the two defining properties of the
Lotka-Volterra model for competing species: \emph{Autocatalysis} (the
synthesis of the peripheral enzymes for both substrates is autocatalytic),
and \emph{mutual inhibition} (the peripheral enzymes for each substrate
inhibit the synthesis of the peripheral enzymes for the other substrate).
The existence of this similarity implies that the dynamics of the
peripheral enzymes are analogous to the dynamics of the Lotka-Volterra
model. In particular, the prediction of diauxic growth by these models
corresponds to {}``extinction'' of one of the enzymes.
\item The models differ with respect to the predicted range of dynamics
and the mechanism by which they inherit the essential properties of
the Lotka-Volterra model.

\begin{enumerate}
\item The B-model captures only diauxic growth patterns, whereas the N-
and T-models capture both diauxic non-diauxic growth patterns
\item In the N-model, mutual inhibition arises because each enzyme stimulates
the dilution of the other enzyme. On the other hand, in the T-model,
the mutual inhibition occurs because the sugar-specific enzymes of
the phosphotransferase system compete for phosphoryl groups.
\end{enumerate}
\end{enumerate}
Comparison with experiments suggests that elements of both all the
models are required for capturing the data.

\section{The models}

Before describing the models, it is useful to mention a few points.

\begin{enumerate}
\item Although all models contain more or less the same variables, the notation
varies considerably from one study to another. To facilitate comparison
between the models, we have used the same notation for the variables.
We denote the cells, exogenous substrates, inducers, and peripheral
enzymes by $C$, $S_{i}$, $X_{i}$ and $E_{i}$, respectively. The
concentrations of these entities are denoted by the corresponding
lower-case letters, $s_{i}$, $e_{i}$, $x_{i}$ and $c$, respectively.
\item All the models assume the existence of a small \emph{constitutive}
or background enzyme synthesis rate that persists even in the absence
of the inducer. We neglect this term since it is generally small compared
to the \emph{induced} enzyme synthesis rate.
\item The cell density and exogenous substrate concentrations are based
on the volume of the culture (gdw/L and g/L, respectively). In contrast,
the concentrations of intracellular variables, such as the enzymes
and inducers, are based on the dry weight of the cells (g/gdw).
\item The foregoing choice of units implies that if $Z$ is any intracellular
entity produced at the rate, $r_{z}^{+}$~g/gdw-hr, and degraded
at the rate, $r_{z}^{-}$~g/gdw-hr, then the mass balance for~$z$
(in g/gdw) is given by the equation\[
\frac{d(zc)}{dt}=\left(r_{z}^{+}-r_{z}^{-}\right)c\Rightarrow\frac{dz}{dt}=r_{z}^{+}-r_{z}^{-}-\left(\frac{1}{c}\frac{dc}{dt}\right)z.\]
Here, the last term reflects the dilution of $Z$ due to growth.
\end{enumerate}
We are now ready to describe the key features of the models.

\begin{figure}
\begin{center}\subfigure[]{\includegraphics[%
  width=7cm,
  height=5cm]{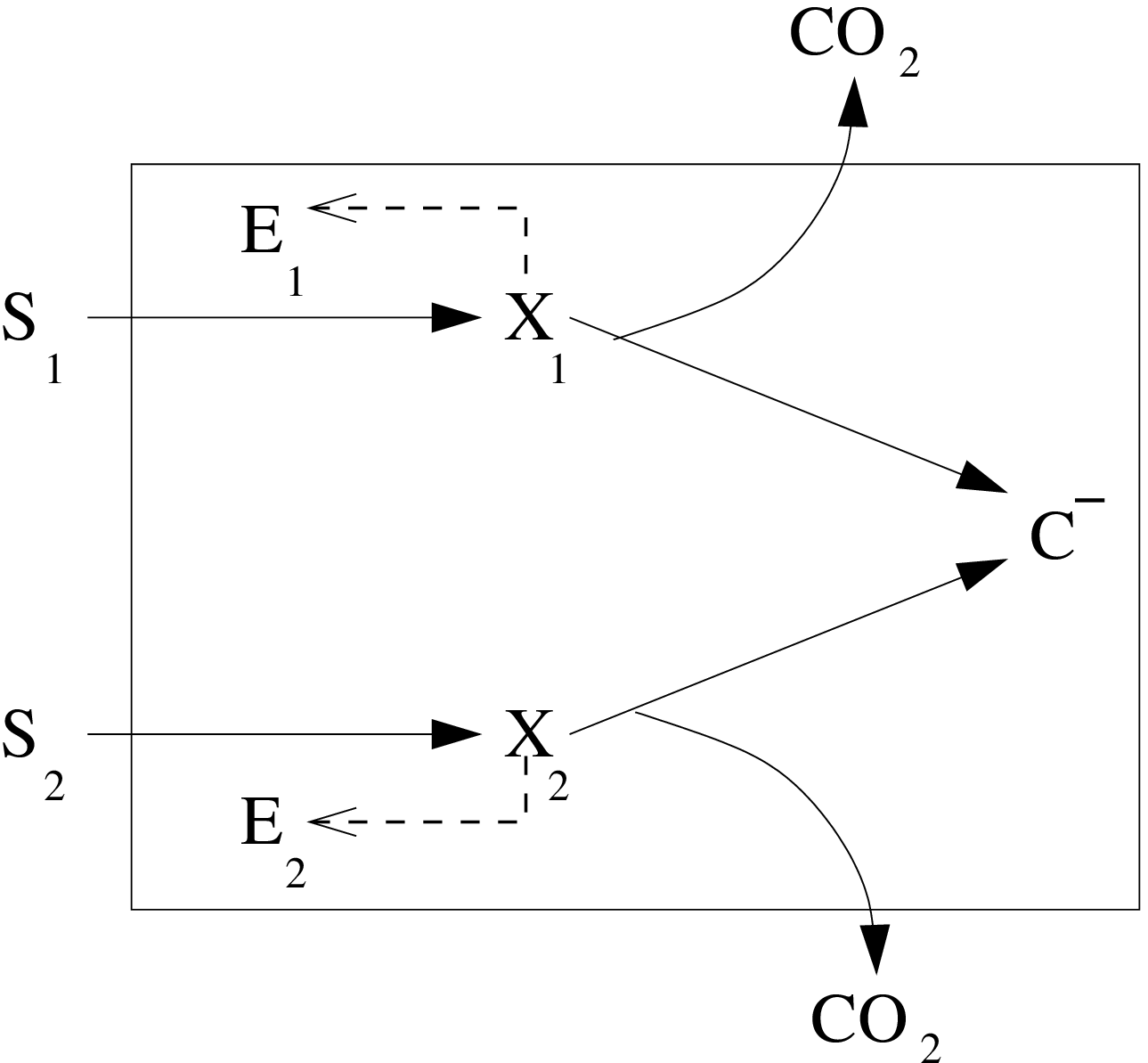}}\hspace{0.3in}\subfigure[]{\includegraphics[%
  width=7cm,
  height=5cm]{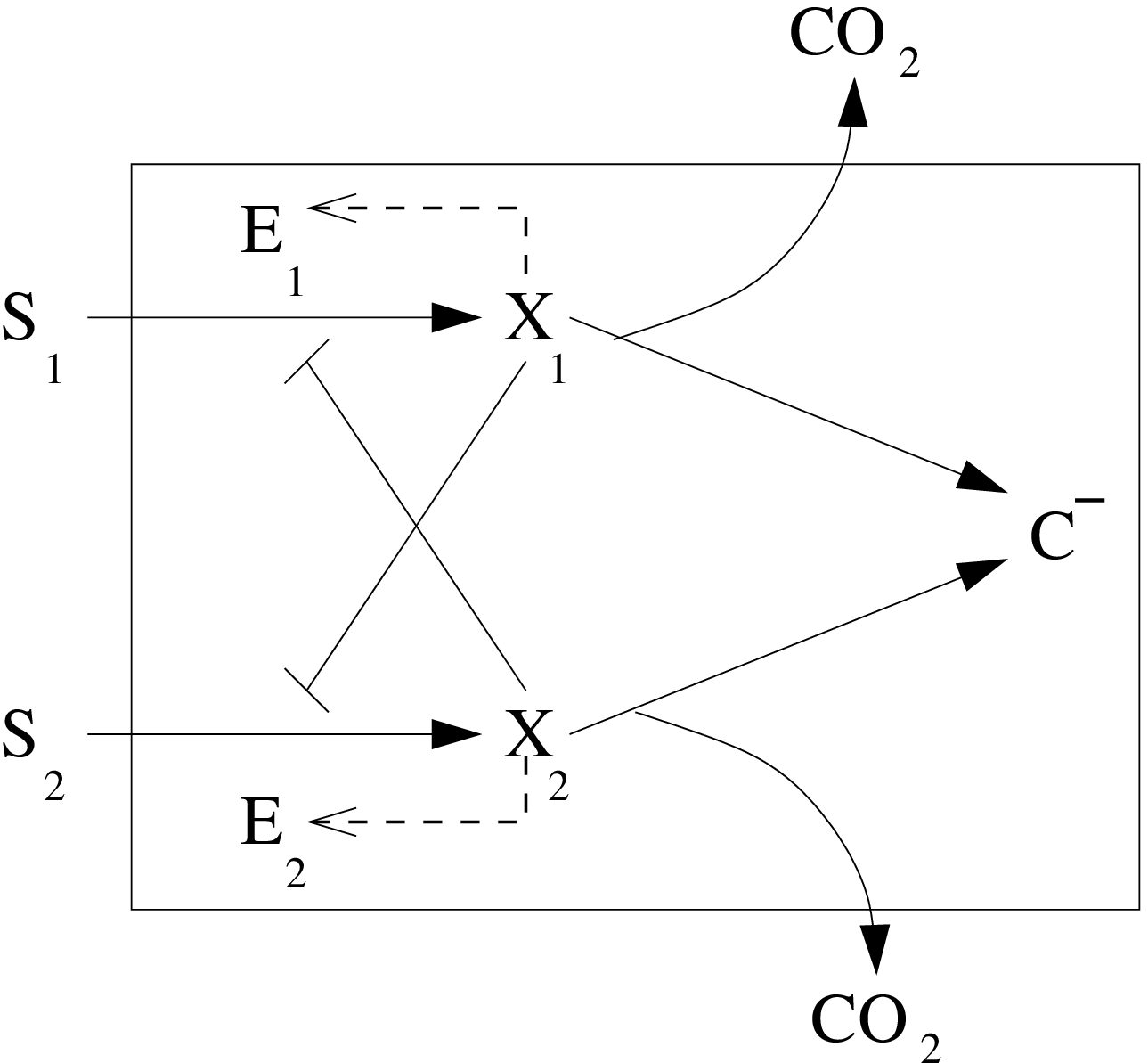}}\end{center}

\caption{\label{f:ModelSchemes1}Kinetic schemes: (a) N-model~\citep{kompala86}.
(b) B-model~\citep{narang98b}.}
\end{figure}

\subsection{N-model}

The kinetic scheme for the N-model is shown in Figure~\ref{f:ModelSchemes1}a.
It is assumed that~\citep{narang98b}

\begin{enumerate}
\item Transport of $S_{i}$ is catalyzed by enzyme $E_{i}$. The specific
uptake rate of $S_{i}$, denoted $r_{s,i}$, follows the kinetics,
$r_{s,i}\equiv V_{s,i}e_{i}s_{i}/(K_{s,i}+s_{i})$.
\item Part of the internalized substrate, denoted $X_{i}$, is converted
to biosynthetic constituents such as amino acids and proteins, denoted
$C^{-}$. The remainder is oxidized to energy (${\rm CO_{2}}$).

\begin{enumerate}
\item The specific rate of conversion of $X_{i}$ to $C^{-}$ and ${\rm CO_{2}}$
is $r_{x,i}\equiv k_{x,i}x_{i}$.
\item The fraction of $X_{i}$ converted to $C^{-}$ is a constant (parameter),
denoted $Y_{i}$. Thus, the specific rate of biosynthesis from $X_{i}$
is $Y_{i}r_{x,i}$.
\end{enumerate}
\item The internalized substrate induces the synthesis of $E_{i}$.

\begin{enumerate}
\item The specific synthesis rate of $E_{i}$ is $r_{e,i}\equiv V_{e,i}x_{i}^{n_{i}}/(K_{e,i}^{n_{i}}+x_{i}^{n_{i}})$,
where $n_{i}=1\textnormal{ or }2$.%
\footnote{Enzyme induction can be hyperbolic ($n_{i}=1$) or sigmoidal ($n_{i}=2$),
depending on the number of inducer molecules that bind to a repressor
molecule~\citep{chung96,yagil71}.%
}
\item The synthesis of the enzymes occurs at the expense of the biosynthetic
constituents, $C^{-}$.
\end{enumerate}
\end{enumerate}
Thus, one obtains the equations\begin{align}
\frac{ds_{i}}{dt} & =-r_{s,i}c,\; r_{s,i}\equiv V_{s,i}e_{i}\frac{s_{i}}{K_{s,i}+s_{i}},\label{eq:NsO}\\
\frac{dx_{i}}{dt} & =r_{s,i}-r_{x,i}-\left(\frac{1}{c}\frac{dc}{dt}\right)x_{i},r_{x,i}\equiv k_{x,i}x_{i},\label{eq:NxO}\\
\frac{de_{i}}{dt} & =r_{e,i}-\left(\frac{1}{c}\frac{dc}{dt}\right)e_{i},\; r_{e,i}\equiv V_{e,i}\frac{x^{n_{i}}}{K_{e,i}^{n_{i}}+x^{n_{i}}},\label{eq:NeO}\\
\frac{dc^{-}}{dt} & =(Y_{1}r_{x,1}+Y_{2}r_{x,2})-(r_{e,1}+r_{e,2})-\left(\frac{1}{c}\frac{dc}{dt}\right)c^{-}.\label{eq:NcMO}\end{align}
These equations implicitly define the specific growth rate and the
evolution of the cell density. To see this, observe that since all
the intracellular concentrations are expressed as mass fractions (g/gdw),
their sum equals~1, i.e., $x_{1}+x_{2}+e_{1}+e_{2}+c^{-}=1$. Hence,
addition of equations (\ref{eq:NxO}--\ref{eq:NcMO}) yields\[
0=\sum_{i=1}^{2}r_{s,i}-(1-Y_{i})r_{x,i}-\frac{1}{c}\frac{dc}{dt}\]
which can be rewritten in the more familiar form\begin{equation}
\frac{dc}{dt}=r_{g}c,\; r_{g}\equiv\sum_{i=1}^{2}r_{s,i}-(1-Y_{i})r_{x,i}\label{eq:NrG}\end{equation}
where $r_{g}$ denotes the specific growth rate.

We can simplify the model by observing that $x_{i}\sim10^{-3}$~g/gdw~\citep{chung96}
and $r_{s,i},r_{x,i}\sim1$~g/gdw-hr. Thus, $x_{i}$ attains quasisteady
state on a time scale of $10^{-3}$~hr. Moreover, the dilution term
$r_{g}x_{i}\sim10^{-3}$~g/gdw-hr is negligibly small compared to
$r_{s,i},r_{x,i}$. Hence, within a few seconds, (\ref{eq:NxO}) becomes,
$0\approx r_{s,i}-r_{x,i}$, so that $r_{g}\equiv\sum_{i}r_{s,i}-(1-Y_{i})r_{x,i}\approx\sum_{i}Y_{i}r_{s,i}$.
Thus, we arrive at the equations \begin{align}
\frac{dc}{dt} & =(Y_{1}r_{s,1}+Y_{2}r_{s,2})c,\; r_{s,i}\equiv V_{s,i}e_{i}\frac{s_{i}}{K_{s,i}+s_{i}}\label{eq:Nc}\\
\frac{ds_{i}}{dt} & =-r_{s,i}c\label{eq:Ns}\\
\frac{de_{i}}{dt} & =r_{e,i}-(Y_{1}r_{s,1}+Y_{2}r_{s,2})e_{i},\; r_{e,i}\equiv V_{e,i}\frac{x_{i}^{n_{i}}}{K_{e,i}^{n_{i}}+x_{i}^{n_{i}}}\label{eq:Ne}\\
x_{i} & \approx\frac{V_{s,i}e_{i}s_{i}/(K_{s,i}+s_{i})}{k_{x,i}}\label{eq:Nx}\\
c^{-} & =1-x_{1}-x_{2}-e_{1}-e_{2}\label{eq:NcM}\end{align}
where (\ref{eq:Nx}) is obtained by solving the quasisteady state
relation, $r_{x,i}\approx r_{s,i}$, for $x_{i}$. Substituting (\ref{eq:Nx})
in the expression for $r_{e,i}$ yields\[
r_{e,i}=V_{e,i}\frac{[e_{i}s_{i}/(K_{s,i}+s_{i})]^{n_{i}}}{\bar{K}_{e,i}^{n_{i}}+[e_{i}s_{i}/(K_{s,i}+s_{i})]^{n_{i}}},\;\bar{K}_{e,i}\equiv K_{e,i}\frac{k_{x,i}}{V_{s,i}}\]
 which shows that enzyme synthesis is autocatalytic: The larger the
enzyme level, the higher its synthesis rate. This is a consequence
of the cyclic structure associated with the kinetics of induction.
Figure~\ref{f:ModelSchemes1}b shows that the enzyme, $E_{i}$, promotes
the synthesis of the inducer, $X_{i}$, which in turn stimulates the
synthesis of even more $E_{i}$. This cycle of reactions implies that
enzyme synthesis is autocatalytic.

\subsection{B-model}

The B-model is similar to the N-model, the only difference being that
the intracellular substrate, $X_{i}$, not only stimulates the induction
of $E_{i}$, but also inhibits the induction of $E_{j},j\ne i$ (shown
in Figure~\ref{f:ModelSchemes2}b as arrows with a bar at one end).%
\footnote{Brandt et al refer to the intracellular substrate (inducer) and the
enzyme induction machinery as \emph{signal molecule} and \emph{synthesizing
unit} (SU), respectively (see Figure~1 of~\citep{Brandt2003}).%
} Assuming that $x_{i}$ rapidly attains quasisteady state, Brandt
et al arrive at the equations~\citep{Brandt2003} \begin{align}
\frac{dc}{dt} & =r_{g}c,\; r_{g}=r_{g,1}+r_{g,2},\; r_{g,i}\equiv V_{g,i}e_{i}\frac{s_{i}}{K_{s,i}+s_{i}}\label{eq:Bc}\\
\frac{ds_{i}}{dt} & =-r_{s,i}c,\; r_{s,i}=\frac{r_{g,i}}{Y_{i}}\label{eq:Bs}\\
\frac{de_{i}}{dt} & =r_{e,i}-r_{g}e_{i},\; r_{e,i}\equiv r_{g}\left[\frac{p_{i}e_{i}s_{i}/(K_{s,i}+s_{i})}{p_{1}e_{1}s_{1}/(K_{s,1}+s_{1})+p_{2}e_{2}s_{2}/(K_{s,2}+s_{2})}\right]\label{eq:Be}\end{align}
where $r_{g,i},r_{s,i},Y_{i}$ denote the specific growth rate, the
specific substrate uptake rate, and the yield of biomass on the $i^{{\rm th}}$
substrate, $r_{g}$ denotes the total specific growth rate, and $0\le p_{i}\le1$
are parameters called \emph{substrate preference coefficients}.

Evidently, enzyme synthesis is autocatalytic because of the positive
feedback from $X_{i}$ to $E_{i}$. The appearance of the specific
growth rate, $r_{g}$, in the expression for $r_{e,i}$ stems from
an additional assumption. It is argued that the specific enzyme synthesis
rate should be proportional to the specific growth rate to ensure
that the specific enzyme synthesis rate grows in proportion to the
specific growth rate.

We note finally that Brandt et al scaled the enzyme level, $e_{i}$,
with the level, $e_{i}^{*}$, that would be observed during exponential
growth on $S_{i}$ alone. Thus, the variable $e_{i}$ shown in equation
(\ref{eq:Be}) corresponds to the scaled variable, $\kappa_{i}\equiv e_{i}/e_{i}^{*}$
in~\citep{Brandt2003}.

\subsection{T-model}

\begin{figure}
\begin{center}\subfigure[]{\includegraphics[%
  width=7cm,
  height=3cm]{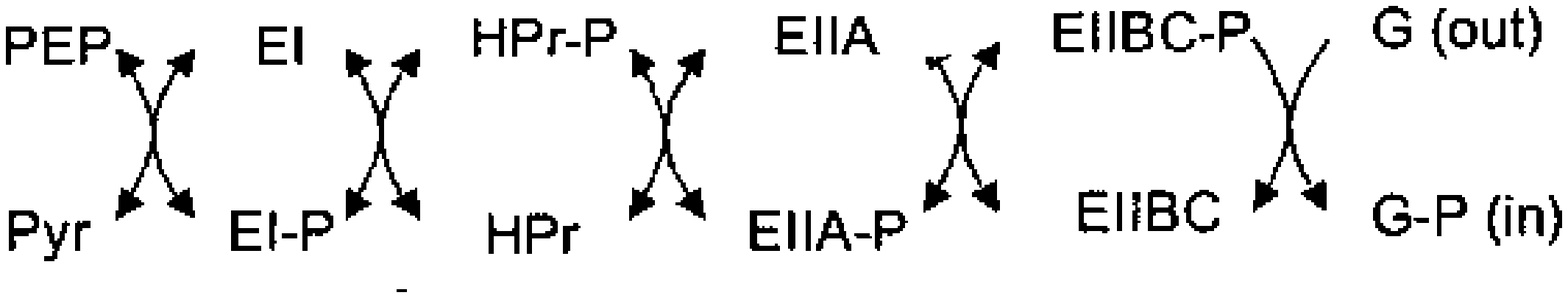}}\hspace*{0.2in}\subfigure[]{\includegraphics[%
  width=7cm,
  height=5cm]{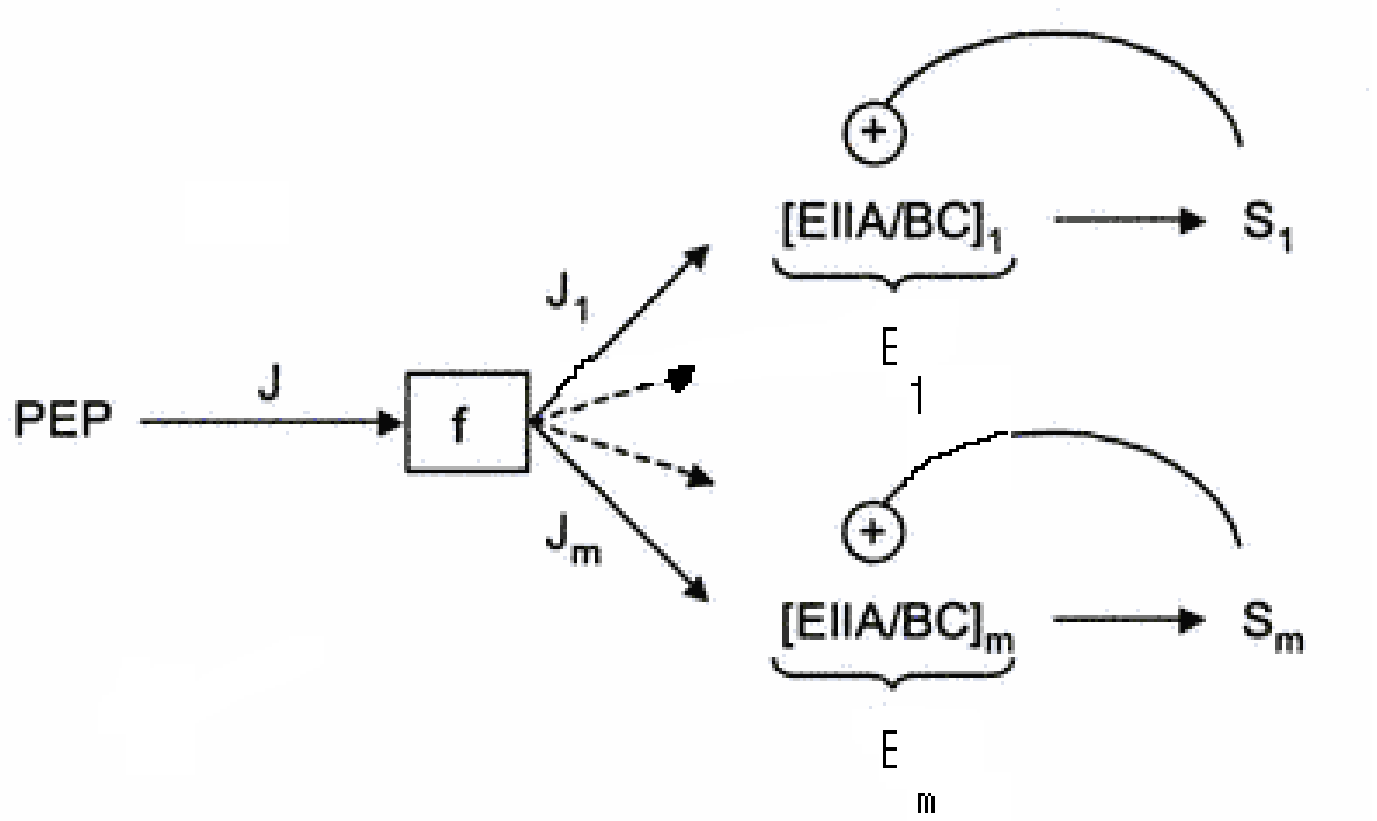}}\end{center}

\caption{\label{f:ModelSchemes2}T-model~\citep{Thattai2003} (a) The phosphotransferase
system (PTS) for carbohydrate uptake. (b) The kinetic scheme.}
\end{figure}

The T-model is aimed at describing the evolution of the peripheral
enzymes belonging to the \emph{phosphotransferase system} (PTS), which
catalyzes the uptake of various sugars in certain bacteria~\citep{Postma1993}.
The uptake of PTS sugars is coupled to their phosphorylation. This
is mediated by a cascade of 5~phosphorylation reactions involving
the successive transfer of a phosphoryl group from PEP to the sugar
(Figure~\ref{f:ModelSchemes2}a). The first two steps, involving
phosphorylation of enzyme~I (EI) and HPr, are common to all the PTS
sugars. The last three steps are mediated by sugar-specific enzymes,
enzyme~IIA (EIIA) and the enzyme~IIBC complex (EIIBC), which ultimately
transfer the phosphoryl group to the sugar during its translocation
across the membrane. The sugar-specific enzymes are inducible, and
their synthesis is coupled since they lie on the same operon.

Figure~\ref{f:ModelSchemes2}b shows the kinetic scheme of the T-model.
It is assumed that

\begin{enumerate}
\item There is a maximum flux, $J$, of phosphoryl groups through the common
enzymes, EI and HPr, and the sugar-specific enzymes of PTS compete
for these phosphoryl group. It turns out that at quasisteady state,
the specific phosphorylation rate of $i^{{\rm th}}$ substrate, $J_{i}$,
(which is equal to the specific substrate uptake rate, $r_{s,i}$)
is given by\[
r_{s,i}=J_{i}=J\frac{\tau_{i}}{1+\tau_{1}+\tau_{2}},\;\tau_{i}\equiv\frac{e_{i}^{2}}{\beta_{i}}\frac{s_{i}}{s_{i}+e_{i}},\]
where $\tau_{i}$ can be interpreted as the \emph{demand} for phosphoryl
groups by the $i^{{\rm th}}$ sugar --- it is an increasing function
of the exogenous sugar concentration, $s_{i}$, and the sugar-specific
enzyme level, $e_{i}$. Consistent with this interpretation, $J_{i}<J$
due to competing demands for phosphoryl groups imposed by the substrates.
\item The specific rate of enzyme synthesis, $r_{e,i}$, is proportional
to the concentration of the intracellular substrate (inducer), $x_{i}$,
which in turn is proportional to the specific substrate uptake rate,
$r_{s,i}$. Hence\[
r_{e,i}\propto r_{s,i},\]
which implies that enzyme synthesis is autocatalytic.
\item The substrate concentrations and the specific growth rate are constant
--- they are treated as parameters.
\end{enumerate}
Thus, the evolution of the sugar-specific enzymes for the $i^{{\rm th}}$
substrate, when appropriately scaled, is given by the equations~\citep{Thattai2003}

\begin{equation}
\frac{de_{i}}{dt}=\frac{\tau_{i}}{1+\tau_{1}+\tau_{2}}-e_{i},\;\tau_{i}\equiv\frac{e_{i}^{2}}{\beta_{i}}\frac{s_{i}}{s_{i}+e_{i}}.\label{eq:Te}\end{equation}
Note that the specific growth rate does not appear in the equations,
since time is scaled by the parameter,~$1/r_{g}$.

\section{\label{s:Results}Results}

We wish to compare the dynamics of the three models described above.
At first sight, this seems impossible since the T-model describes
the dynamics of the enzymes only, whereas the N- and B-models describe
the dynamics of the enzymes, substrates, and cells. It turns out,
however, that the dynamics of the substrates and cells are irrelevant
on the time scale of interest. Indeed, insofar as the dynamics of
mixed-substrate growth are concerned, the asymptotic dynamics ($t\rightarrow\infty$)
of the N- and B-models is of little interest. Much more revealing
are their dynamics during the first exponential growth phase, since
it is these finite-time dynamics that determine the substrate utilization
pattern. Specifically, diauxic growth will occur if the peripheral
enzymes for one of the substrates vanishes during the first exponential
growth phase. In contrast, simultaneous substrate utilization will
be observed if the enzymes for both substrates persist during the
first exponential growth phase. We show below that

\begin{enumerate}
\item In the N-, and B-models, the motion of the enzymes during the first
exponential growth phase can be described by a reduced \emph{}system
of two equations that are formally similar to the equations of the
T-model. This makes it possible to compare the N- and B-models with
the T-model.
\item The reduced equations of all the models are different realizations
of the generalized Lotka-Volterra model for two competing species.
Thus, in all the models, the enzymes behave like two competing species.
In particular, they coexist or become extinct, and these dynamics
have meaningful biological interpretations in the context of mixed-substrate
growth.
\item The B-model can never capture non-diauxic growth patterns.
\end{enumerate}
Finally, we compare the mechanisms underlying the dynamics of the
N- and T-models.

\subsection{All the models are dynamically similar to the Lotka-Volterra model}

We begin by showing that in the N- and B-models, the dynamics of the
enzymes during the first exponential growth phase can be described
by a reduced system of two equations. To see this, observe that during
this period, both substrates are in excess, i.e., $s_{i}\gg K_{s,i}$.
Hence, even though the exogenous substrate concentrations are changing,
the transport enzymes remain saturated ($s_{i}/(K_{s,i}+s_{i})\approx$1).
Now, the cells sense the environment through the transport enzymes.
Since these enzymes see a quasiconstant environment during the first
exponential growth phase, they approach quasisteady state levels.
It follows that in the N- and B-models, the motion of the enzymes
from any initial conditions to the quasisteady state levels can be
obtained from~(\ref{eq:Ne}) and~(\ref{eq:Be}) by replacing $s_{i}/(K_{s,i}+s_{i})$
with~1.%
\footnote{We have reduced the equations by appealing to intuitive arguments.
This reduction can be justified rigorously by appealing to the theorem
of continuous dependence on initial conditions (see~\citep{narang97c}
for details). %
} Thus, we arrive at the reduced equations\begin{align}
\frac{de_{i}}{dt} & =V_{e,i}\frac{e_{i}^{n_{i}}}{K_{e,i}+e_{i}^{n_{i}}}-\left(Y_{1}V_{s,1}e_{1}+Y_{2}V_{s,2}e_{2}\right)e_{i}\label{eq:NeR}\\
\frac{de_{i}}{dt} & =r_{g}\left(\frac{p_{i}e_{i}}{p_{1}e_{1}+p_{2}e_{2}}-e_{i}\right).\label{eq:BeR}\end{align}
Since these equations are formally similar to equation~(\ref{eq:Te}),
we can compare the dynamics of all three models.

It turns out that the equations of all three models are dynamical
analogs of the generalized Lotka-Volterra model for two competing
species, which is given by the equations~\citep[Chapter 12]{hirsch}\begin{equation}
\frac{dN_{i}}{dt}=f_{i}(N_{1},N_{2})N_{i}\label{eq:GLK}\end{equation}
where $N_{i}$ and $f_{i}(N_{1},N_{2})$ denote the population density
and specific growth rate of the $i^{{\rm th}}$ species, respectively,
and $f_{i}(N_{1},N_{2})$ satisfies the properties

\begin{enumerate}
\item $\partial f_{1}/\partial N_{2},\partial f_{2}/\partial N_{1}<0$,
i.e., each species inhibits the growth of the other species.
\item $f_{i}(N_{1},N_{2})<0$ for sufficiently large $N_{1},N_{2}>0$, i.e.,
at sufficiently large population densities, the specific growth rates
are negative.
\end{enumerate}
The standard Lotka-Volterra model for competing species is a special
case of the generalized model with \[
f_{i}(N_{1},N_{2})=r_{i}(1-a_{i1}N_{1}-a_{i2}N_{2})\]
where $r_{i}$ is the unrestricted specific growth rate of the $i^{{\rm th}}$
species in the absence of any competition, and $a_{i1},a_{i2}$ are
coefficients that quantify the reduction of the unrestricted specific
growth rate due to intra- and inter-specific competition~\citep{murray}.
The analogy between the generalized Lotka-Volterra model and equations
(\ref{eq:Te}--\ref{eq:BeR}) becomes evident if we rewrite the latter
in the form\begin{align}
\frac{de_{i}}{dt} & =f_{i}^{N}(e_{1},e_{2})e_{i},\; f_{i}^{N}(e_{1},e_{2})\equiv V_{e,i}\frac{e_{i}^{n_{i}-1}}{\bar{K}_{e,i}^{n_{i}}+e_{i}^{n_{i}}}-\left(Y_{1}V_{s,1}e_{1}+Y_{2}V_{s,2}e_{2}\right)\label{eq:NeR1}\\
\frac{de_{i}}{dt} & =f_{i}^{B}(e_{1},e_{2})e_{i},\; f_{i}^{B}(e_{1},e_{2})\equiv r_{g}\left(\frac{p_{i}}{p_{1}e_{1}+p_{2}e_{2}}-1\right)\label{eq:BeR1}\\
\frac{de_{i}}{dt} & =f_{i}^{T}(e_{1},e_{2})e_{i},\; f_{i}^{T}(e_{1},e_{2})\equiv\frac{e_{i}s_{i}/(\beta_{i}+e_{i})}{1+\tau_{1}+\tau_{2}}-e_{i},\;\tau_{i}\equiv\frac{e_{i}^{2}}{\beta_{i}}\frac{s_{i}}{s_{i}+e_{i}}\label{eq:TeR1}\end{align}
One can check that the functions, $f_{i}^{N},f_{i}^{B},f_{i}^{T}$
satisfy the properties 1 and 2 above. Thus, in all the models, the
dynamics of the enzymes during the first exponential growth phase
are analogous to the dynamics of the generalized Lotka-Volterra model
for two competing species.

The dynamics of the generalized Lotka-Volterra model for competing
species are well understood~\citep[Chapter 12]{hirsch}. Specifically,
the model entertains no limit cycles, so that all solutions ultimately
converge to some steady state. Despite the absence of limit cycles,
the model has a rich spectrum of dynamics. Even in the case of the
standard Lotka-Volterra model, one can get 4~different types of dynamics
depending on the parameter values~\citep{murray}. Indeed, if we
define the dimensionless variables, $u_{i}\equiv a_{ii}N_{i}$ and
$\tau\equiv r_{1}t$, we obtain the dimensionless equations \begin{align*}
\frac{du_{1}}{d\tau} & =(1-u_{1}-b_{12}u_{2})u_{1},\; b_{12}\equiv\frac{a_{12}}{a_{11}}\\
\frac{du_{2}}{d\tau} & =\rho(1-b_{21}u_{1}-u_{2})u_{2},\;\rho\equiv\frac{r_{2}}{r_{1}},\; b_{21}\equiv\frac{a_{21}}{a_{22}}.\end{align*}
The steady states of the model are completely determined by the parameters,
$b_{ij}$, which may be viewed as a measure of the extent to which
the $j^{{\rm th}}$ species inhibits the $i^{{\rm th}}$ species.
Figure~\ref{f:GlobalDynamics}a shows the bifurcation diagram of
the scaled standard Lotka-Volterra model. The bifurcation diagram
shows that when neither species inhibits the other species strongly
($b_{12},b_{21}<1$), the two species coexist; when the cross-inhibition
is asymmetric ($b_{12}>1,b_{21}<1$ or $b_{12}<1,b_{21}>1$), one
of the species is rendered extinct; when both species inhibit each
other strongly ($b_{12},b_{21}>1$), the outcome of the competition
depends on the initial condition.

Given the dynamical analogy between the Lotka-Volterra model and equations~(\ref{eq:NeR1}--\ref{eq:TeR1}),
it is reasonable to expect that the peripheral enzymes would yield
similar dynamics during the first phase of exponential growth. Importantly,
these dynamics have simple interpretations in terms of the substrate
utilization pattern. Indeed, extinction of one of the enzymes during
the first phase of exponential growth corresponds to diauxic growth;
coexistence of the enzymes during this period is the correlate of
simultaneous substrate uptake; and bistability reflects a substrate
utilization pattern which varies depending on the manner in which
the inoculum has been precultured.

\subsection{The B-model cannot capture non-diauxic growth patterns}

\begin{figure}
\begin{center}\subfigure[]{\includegraphics[%
  width=7cm,
  height=7.5cm]{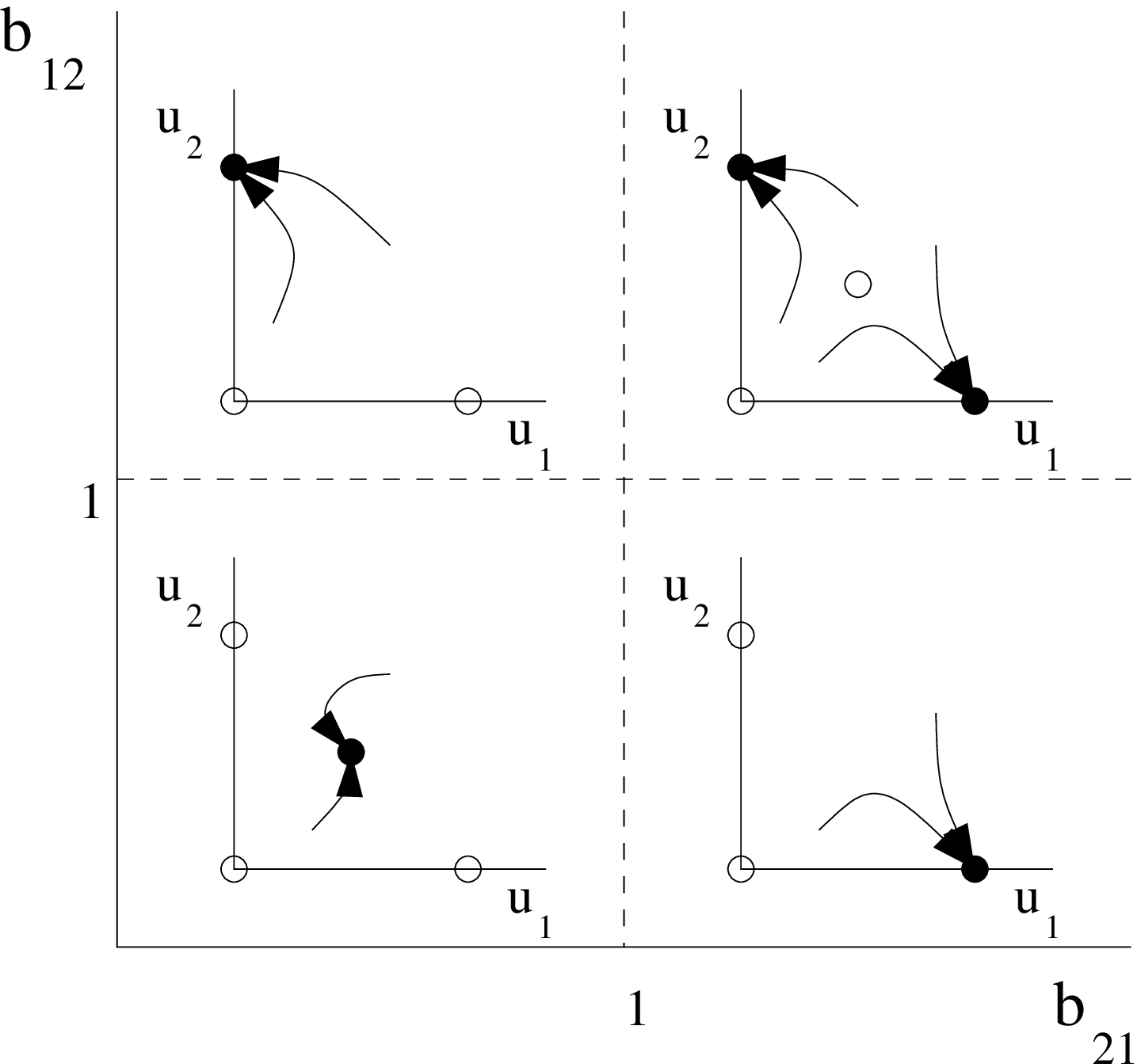}}\hspace{0.3in}\subfigure[]{\includegraphics[%
  width=7cm,
  keepaspectratio]{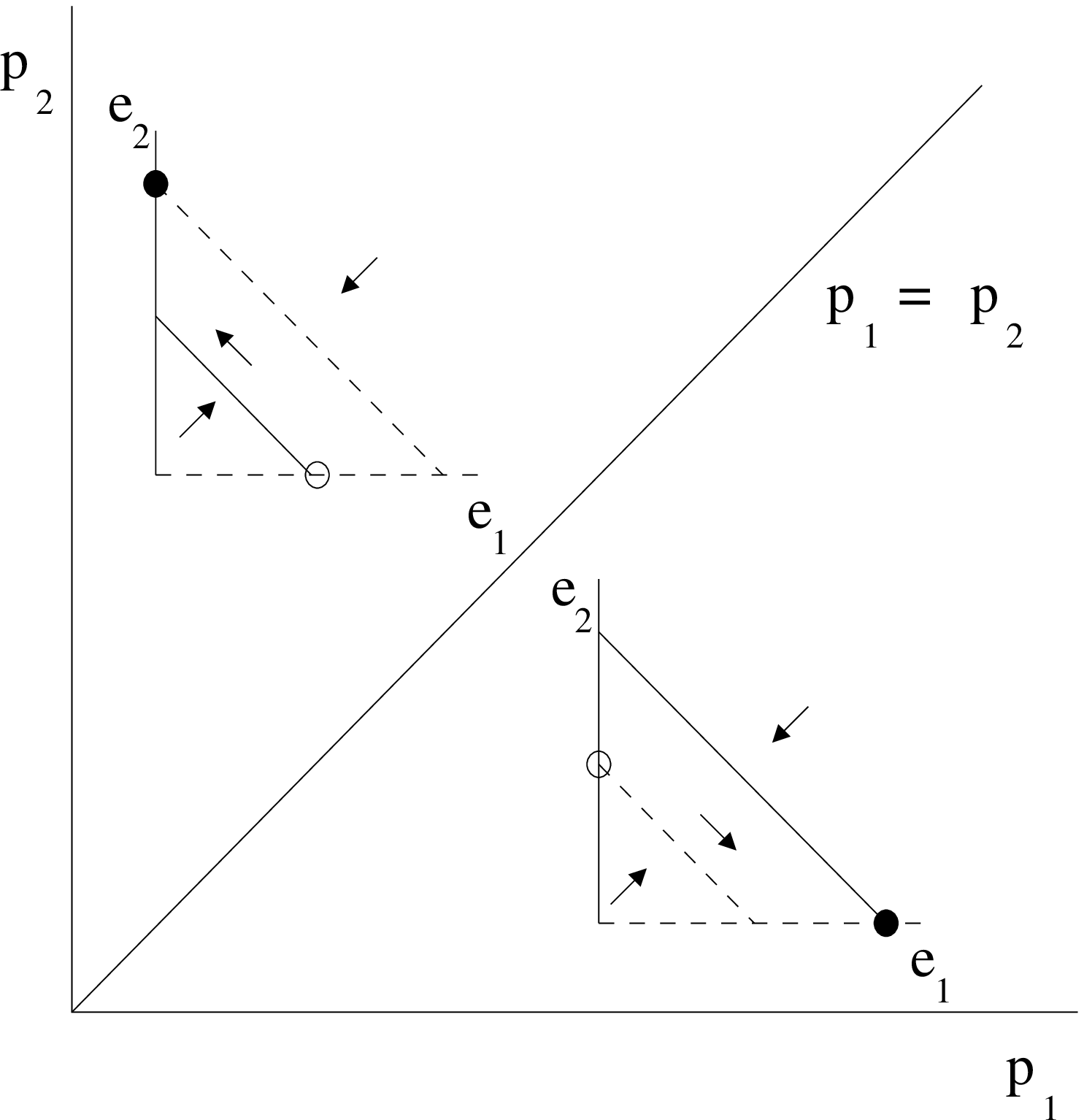}}\end{center}

\caption{\label{f:GlobalDynamics}Classification of the global dynamics: (a)
The bifurcation diagram of the standard Lotka-Volterra model. The
full and open circles show stable and unstable steady states, respectively.
(b) The bifurcation diagram of the B-model. When $p_{1}>p_{2}$ (resp.,
$p_{2}>p_{1}$), $E_{2}$ (resp., $E_{1}$) is rendered extinct during
the first exponential growth phase. The full and dashed lines show
the nullclines for $e_{1}$ and $e_{2}$, respectively. The full and
open circles show stable and unstable steady states, respectively.
The arrows show the orientation of the vector, $(de_{1}/dt,de_{2}/dt)$,
in the regions between the nullclines.}
\end{figure}
The N- and T-models can capture diauxic growth, simultaneous substrate
utilization, and bistable growth (see~\citep{narang98b,Thattai2003}
for details). However, the B-model always exhibits diauxic growth.
Indeed, the dynamics are completely determined by the substrate preference
coefficients, $p_{1}$ and $p_{2}$ (Figure~\ref{f:GlobalDynamics}b).
If $p_{1}>p_{2}$, $E_{2}$ becomes extinct during the first exponential
growth phase, which corresponds to preferential consumption of $S_{1}$.
Conversely, if $p_{2}>p_{1}$, $E_{1}$ becomes extinct during the
first exponential growth phase, which corresponds to preferential
consumption of $S_{2}$.

To see this, it suffices to consider the nullclines of equation~(\ref{eq:BeR1}),
i.e., the curves along which $de_{i}/dt=0$. These curves, which separate
the $e_{1}e_{2}$-plane into regions in which $de_{i}/dt$ is nonzero,
are given by the equations\[
e_{i}=0\textrm{ or }f_{i}^{B}(e_{1},e_{2})\equiv\frac{p_{i}}{p_{1}e_{1}+p_{2}e_{2}}-1=0,\; i=1,2.\]
Evidently, $de_{1}/dt=0$ along the $e_{2}$-axis and the straight
line, $p_{1}e_{1}+p_{2}e_{2}=p_{1}$; we shall refer to the latter
curve as $\mu$. Likewise, $de_{2}/dt=0$ along the $e_{1}$-axis
and the straight line, $p_{1}e_{1}+p_{2}e_{2}=p_{2}$; we shall refer
to the latter curve as $\nu$. The steady states of (\ref{eq:BeR1})
lie at the intersection points of the nullclines for $e_{1}$ and
$e_{2}$. Since $\mu$ and $\nu$ are parallel, there are no {}``coexistence''
steady states. There are {}``extinction'' steady states at$(1,0)$
and $(0,1)$.%
\footnote{There is no steady state at $(0,0)$ since the model is undefined
(discontinuous) at this point.%
} One can check that the stability of $(1,0)$ and $(0,1)$ is determined
by the disposition of $\mu$ and $\nu$. If $p_{1}>p_{2}$, then $\mu$
lies above $\nu$, and $(1,0)$ is stable, while $(0,1)$ is unstable.
Conversely, if $p_{1}<p_{2}$, then $\nu$ lies above $\mu$, and
$(0,1)$ is stable, while $(1,0)$ is unstable. Thus, we conclude
that the B-model entertains only the diauxic growth pattern.

\subsection{The N- and T-models have different mechanisms of mutual inhibition}

We have shown above that all the models are different realizations
of the generalized Lotka-Volterra model for two competing species.
Furthermore, the B-model cannot capture the non-diauxic growth patterns.
In what follows, we consider the similarities and differences between
the N- and T-models.

We can develop a better appreciation of the similarities and differences
by examining the manner in which these models acquire the properties
of the generalized Lotka-Volterra model. The latter is characterized
by two essential properties.

\begin{enumerate}
\item The growth of each species is autocatalytic, i.e., $dN_{i}/dt=0$
whenever $N_{i}=0$.
\item The interaction between the two species is \emph{mutually inhibitory},
i.e, $\partial f_{1}/\partial N_{2},\partial f_{2}/\partial N_{1}<0$.
\end{enumerate}
It is clear that all the models satisfy the first property precisely
because enzyme synthesis is autocatalytic ($r_{e,i}=0$ whenever $e_{i}=0$).
The mechanism that ensures that existence of this property is also
identical in all the models. It stems from the fact that the enzyme
promotes the formation of the internalized substrate (inducer) which
in turn stimulates the synthesis of even more enzyme.

The difference between the models lies the mechanism(s) leading to
the second property, namely, mutual inhibition. In the N-model, each
enzyme inhibits the other enzyme by stimulating growth, and thus increases
the rate of \emph{dilution} of the other enzyme (e.g., $\partial f_{1}^{N}/\partial e_{2}<0$
precisely because $e_{2}$ appears in the dilution term for $e_{1}$).
On the other hand, in the T-model, there is no mutual inhibition due
to dilution --- in fact, the specific growth rate is assumed to be
a constant parameter. Instead, each enzyme inhibits the rate of \emph{synthesis}
of the other enzyme (e.g., $\partial f_{1}^{T}/\partial e_{2}<0$
precisely because $e_{2}$ appears in the synthesis term for $e_{1}$),
and this inhibition occurs due to competition for the phosphoryl groups.

The N-model has two advantages over the T-model.

\begin{enumerate}
\item It is more general than the T-model since it applies to any pair of
inducible substrates, as opposed to PTS sugars only. Indeed, the N-model
appeals to the two processes --- enzyme induction and growth --- that
occur in every system involving inducible substrates.
\item It explains an important empirical correlation observed in mixed-substrate
growth. Based on a comprehensive review of the experimental literature,
Harder \& Dijkhuizen~\citep{harder82} and Egli~\citep{egli95}
have observed that in general, both substrates are consumed simultaneously
when they support low-to-medium single-substrate growth rates. On
the other hand, diauxic growth is typically observed when one of the
substrates supports a much higher specific growth rate. In this case,
the substrate supporting the higher specific growth rate is usually
the {}``preferred substrate.'' \\
This can be understood in terms of the N-model, wherein each enzyme
inhibits the other enzyme by enhancing the latter's dilution rate.
Thus, enzymes for two substrates that support low-to-medium growth
rates will coexist since they will not inhibit each other significantly.
However, if the two substrates, say $S_{1}$ and $S_{2}$, support
high and low specific growth rates, respectively, then $E_{1}$ will
strongly inhibit the synthesis of $E_{2}$, but $E_{2}$ will have
little inhibitory effect on synthesis of $E_{1}$. Consequently, $E_{1}$
will drive $E_{2}$ to {}``extinction,'' resulting in preferential
utilization of $S_{1}$.
\end{enumerate}
The disadvantage of the N-model is that, unlike the T-model, it does
not account for inhibition of enzyme synthesis. It is conceivable
that this occurs by competition for phosphoryl groups. Another mechanism,
well-documented in the experimental literature, is inducer exclusion~\citep{stulke99}.
The latter is not accounted for by the N- and T-models. Indeed, neither
one of these models accounts for direct interaction between the enzymes
for the two substrates. The effect of the enzymes belonging to the
other substrate are exerted indirectly by influencing the specific
growth rate or demand for the phosphoryl groups. Thus, the N-model
can be viewed as a general model which is true of every pair of substrates
with inducible peripheral enzymes. However, for quantitative agreement,
it must be modified along the lines of the T-model by accounting for
specific mechanisms, such as inducer exclusion and competition for
phosphoryl groups, that inhibit enzyme synthesis.

It is striking that all the models can predict diauxic growth despite
the absence of direct inhibitory interactions such as inducer exclusion.
These dynamics occur precisely because enzyme synthesis is autocatalytic
--- it is this property that makes it feasible for enzymes to become
{}``extinct'' during the first exponential growth phase. Thus, the
models imply that diauxic growth would not be observed if autocatalysis
were destroyed. This is consistent with the experimental data. Constitutive
mutants, in which synthesis of lactose-specific enzymes persists even
in absence of the inducer concentration ($\left.r_{e,i}\right|_{e_{i}=0}>0$),
do not display the diauxie~\citep[Figure~6]{inada96}. Similarly,
the glucose-lactose diauxie is not observed if the medium contains
IPTG, an inducer of the \emph{lac} operon that can enter the cell
even in the absence of the lactose permease~\citep[Figure~7]{inada96}.

\section{\label{s:Conclusions}Conclusions}

We compared the similarities and differences between three kinetic
models of mixed-substrate growth. We showed that

\begin{enumerate}
\item In all three models, the dynamics of the peripheral enzymes are formally
similar to the generalized Lotka-Volterra model for competing species.
This similarity occurs because the peripheral enzymes mirror the two
essential properties of the Lotka-Volterra model: (a)~Synthesis of
the peripheral enzymes for both substrates are autocatalytic (b)~The
peripheral enzymes for the two substrates inhibit each other.
\item The model by Brandt et al~\citep{Brandt2003} cannot capture non-diauxic
growth patterns. For all parameter values, the peripheral enzymes
for one of the substrates becomes extinct during the first exponential
growth phase, thus resulting in diauxic growth.
\item The models in~\citep{narang98b} and~\citep{Thattai2003} capture
both diauxic and non-diauxic growth patterns. Both models are identical
with respect to the mechanism that ensures that peripheral enzyme
synthesis is autocatalytic --- the peripheral enzymes promote the
synthesis of the inducer, which in turn stimulates the synthesis of
even more enzyme. However, they differ with respect to the mechanism
that produces mutual inhibition. In the Narang model, the mutual inhibition
occurs because each enzyme stimulates the dilution rate of the other
enzyme. In the Thattai \& Shraiman model, which applies to PTS sugars
only, the mutual inhibition stems from competition for phosphoryl
groups.
\end{enumerate}
\bibliographystyle{plainnat}
\bibliography{compAnalysis}

\end{document}